 \documentclass[nohyper,12pt,letterpaper]{JHEP}
 \usepackage{epsfig}
 
 \def \eqn#1#2{\begin{equation}#2\label{#1}\end{equation}}

 \def\bomega{\bf\Omega}
 \title{$II_{\infty}$ Factors and M-theory in\\ Asymptotically Flat Space}
 \author{T.\,Banks\\
 Department of Physics and SCIPP\\
 University of California, Santa Cruz, CA 95064\\
 E-mail: \email{banks@scipp.ucsc.edu}\\
 {\it and}\\
 Department of Physics and NHETC, Rutgers University\\
 Piscataway, NJ 08540}

 \abstract{I discuss a formulation of M-theory at null infinity,
which is based on general principles of holographic space-time, and
is manifestly covariant. The construction utilizes a certain Type II
Von Neumann algebra, which provides a kinematic framework,
alternative to Fock Space, for describing the scattering states of
eleven dimensional asymptotically flat M-theory.  The construction
provides a greatly clarified statement of the connection between
SUSY and holography. I make preliminary remarks about dynamical
equations for the S-matrix, and compactifications.}

 \received{} \accepted{}
 \preprint{hep-th{}\\\\RU-06-08, SCIPP-06-07 \\}
 
\begin{document}

\section{\bf Introduction}

In a series of papers primarily devoted to
cosmology\cite{holcosm}, W. Fischler and the present author have
developed a general framework for quantum gravity based on the
holographic principle.  The purpose of the present paper is to
begin to make contact between that formalism and conventional
formulations of M-theory.   We will deal primarily with the
simplest version of the theory in 11 asymptotically flat
dimensions.

The basic ingredient of the holographic approach to quantum gravity
is the operator algebra of a causal diamond\footnote{A causal
diamond is the region bounded by the backward lightcone of a point P
and the forward lightcone of a point Q in the causal past of P.}.
This replaces the notion of the fields at a point in local field
theory.   The covariant entropy bound \cite{fsb} bounds the entropy
in a causal diamond by one quarter of the area in Planck units of
its holographic screen.  The holographic screen is the maximal area
spacelike $d -2$ surface on the boundary of the diamond.  Fischler
and I hypothesized that the entropy referred to in the entropy bound
was that of the maximally uncertain density matrix in the Hilbert
space of states associated with the diamond.   Thus the entropy is
the logarithm of the dimension of this Hilbert space.

The operator algebra of the diamond can be constructed from the
quantization of classical variables specifying the orientation of
its holoscreen.   A small area or pixel on the holoscreen can be
described by giving the direction of the null ray\footnote{There is
an implicit choice of whether the future directed null ray is
entering or leaving the diamond.  In asymptotically flat space one
should associate different variables with incoming and outgoing
rays, which are related by the approximate S-matrix described
below.} penetrating the screen and a $(d - 2)$ dimensional area
element transverse to this null ray.   Both are specified, up to
conformal rescaling, by a solution of the Cartan-Penrose equation:
\eqn{cp}{\bar{\psi} \gamma^{\mu} \psi \gamma_{\mu} \psi = 0.}   The
independent real components of this pure spinor are quantized by the
formula \eqn{acr}{[S_a (n), S_b (n)]_+ = \delta_{ab} .}  The
dimension of the irreducible representation of this algebra
specifies the area of the pixel, by the Bekenstein-Hawking rule.

The quantization condition breaks the conformal invariance of the CP
equation, leaving over only a (local) $Z_2$.   Using this, we can
Klein transform the {\it a priori} commuting variables associated
with different pixels, so that the full operator algebra of the
diamond is \eqn{acr2}{[S_a (n), S_b (m) ]_+ = \delta_{ab}
\delta_{mn} .} This residual $Z_2$ gauge invariance is identified
with $( - 1)^F$.

The proposed quantization rule says that the degrees of freedom
associated with a pixel of the holographic screen are precisely
the spin degrees of freedom of a massless superparticle, which may
be viewed as the quantum degree of freedom which entered (exited)
the diamond via that pixel.   This connection between the
holographic principle and supersymmetry is one of the most
exciting features of our formalism.

Note that, although we will not discuss compactification of eleven
dimensions in this paper, the way to accommodate it is to enlarge
the algebra of pixel generators to included charges associated
with Kaluza-Klein symmetries, their magnetic duals, and wrapped
branes.   These are precisely the features of compact spaces which
are invariant under the various dualities of M-theory.    We will
propose a quasi-geometrical picture only for the non-compact
dimensions of space-time (which are however taken to include de
Sitter space in some cases).

\section{M-theory in Asymptotically Flat Space (AFM-theory) - Light Cone Gauge}

At present, our most complete formulation of M-Theory in
Asymptotically Flat Spacetime is for those cases where Matrix
Theory\cite{bfss} is well defined.  Matrix Theory is the DLCQ of
M-Theory compactified on $T^d$ with $0\leq d \leq 5$ and on K3
\cite{br}.   For $ d \leq 3$ it is described in terms of a
Lagrangian quantum field theory, while for $d = 4$ and K3 one
needs the $(2,0)$ superconformal field theory in six dimensions,
and for $d=5$ the even more mysterious Little String Theory.

The S-matrix of M-Theory is supposed to be obtained as the
$N\rightarrow \infty$ limit of the scattering matrix of these
quantum theories.  All of them have a moduli space on which one
can define a scattering problem (note that the space on which the
various field theories are defined is compact and does not admit
scattering).  Apart from amplitudes protected by
non-renormalization theorems, one does not expect to get covariant
results before taking the large $N$ limit.

It would clearly be useful to have a formulation in which $N$ was
already infinite.  One would then expect to have exact Lorentz
invariance.  Moreover, some of the problems associated with
compactifying more dimensions appear to go away at $N=\infty$. In
this paper I will present some ideas about formulating theories of
asymptotically flat quantum gravity directly at null infinity, in
a manifestly Lorentz invariant manner.. For the present, I will
restrict attention to the uncompactified eleven dimensional
theory.

The first ideas, which evolved into this paper, were developed
several years ago in an attempt to formulate Matrix Theory
directly at $N = \infty$\footnote{This attempt has run into
various problems. I will present it elsewhere, if a solution to
those problems is found.}. The obvious conjecture is that the
variables of Matrix Theory should be replaced by elements of an
infinite dimensional associative algebra. The algebra should be a
complex algebra with involution, and the basic variables are
Hermitian with respect to the involution.
 The form of the Matrix Theory action
requires that the algebra possess a well defined trace. This then
defines an inner product via \eqn{scalprod}{(A,B) = Tr A^{\dagger} B
.}

Using this one can realize the algebra as an algebra of linear
operators acting on an inner product space.  Various theorems in
the literature suggest\cite{Kap} that it will always in fact be a
Von Neumann algebra, that is, a weakly closed \footnote{Weakly
closed means that if for a sequence of operators , $A_n$, $A_n
|\psi> \rightarrow A |\psi > $, for all states $| \psi >$, then
$A$ is a member of the algebra.} algebra of bounded operators in a
Hilbert space. By a celebrated theorem of Von Neumann, a condition
equivalent to weak closure is that the algebra be equal to its own
double commutant in the algebra of all bounded operators on
Hilbert space. A Von Neumann algebra whose center is just the
complex numbers is called a factor.  Every Von Neumann algebra can
be constructed as a direct sum (integral) of factors.

Murray and Von Neumann \cite{mvn} classified factors according to
the possible values that the trace takes on projectors {\it i.e.}
operators satisfying $e_i^2 = e_i$.  Type $I_N$ algebras are the
algebras of all bounded operators in an $N$ dimensional Hilbert
space ($N=\infty$ is included in the list).  The trace of
projectors is always a positive integer.  In Type II algebras the
trace of projectors takes on continuous positive values.  $II_1$
algebras have a maximal value of the trace, that of the unit
operator, which is normalized to $1$.  In Type $II_{\infty}$
algebras the trace of projectors is unbounded.  Every such algebra
is a direct product of a $I_{\infty}$ factor and a Type $II_1$
factor.  In Type III factors the trace of projectors takes on only
the values $0$ and $\infty$.  We will not have further occasion to
discuss them here.

It is obvious that the factors of interest for us are of Type
$II_{\infty}$.   In fact I will suggest that it is a very
particular $II_{\infty}$ factor.  The trace of an element of the
algebra will measure its longitudinal momentum (in Planck units).
That is, for any operator $A$ in our algebra, find the projector
$e_{p^+}$ of maximal trace ($\equiv p^+$) such that $A = e_{p^+} A
e_{p^+} $.   $P^+ (A) $ is the function from the algebra to the
nonnegative real numbers defined by this maximal trace.

One of the invariants that characterizes inequivalent factors of
the same type, is the group of outer automorphisms of the algebra.
An automorphism is an invertible mapping of the algebra into
itself which preserves the algebraic operations (including
Hermitian conjugation) and is continuous in the topology defined
by weak operator convergence.   If $U$ is any unitary element of
the algebra, the mapping $A \rightarrow U^{\dagger} A U$ is an
{\it inner} automorphism. The group of outer automorphisms, $Out
[{\cal A}]$ is the factor group of the full automorphism group by
the normal subgroup of inner automorphisms, (also called the
unitary group of the algebra) . Given any automorphism $\rho$ we
can define a new trace by $Tr_{\rho} [A] = Tr[\rho (A)]$.  For
inner automorphisms this is just the old trace.  For outer
automorphisms it is in principle different, but Murray and Von
Neumann proved the uniqueness of the trace up to a positive
multiplicative factor. Thus $Tr_{\rho} [A] = e^{\lambda_{\rho}} Tr
[A]$ where $\lambda_{\rho}$ is real.  This mapping from the group
$Out [{\cal A}]$ to the group of positive reals, is a
homomorphism, and the image subgroup is an invariant which can be
used to distinguish factors.

There is a particular $II_{\infty}$ factor for which this group is
the group of all positive real numbers, and the homomorphism is an
isomorphism.   Thus, there is a unique outer automorphism that
multiplies the trace by any given positive number.  For this
factor, which is denoted $R_{0,1}$ in the classification of Araki
and Woods, \cite{arwood}, every two projectors with the same
trace, are unitarily equivalent.

The factor $R_{0,1}$ is the product of the algebra of all bounded
operators (the unique $I_{\infty}$ factor) with a special $II_1$
factor first constructed by Murray and Von Neumann.  The
construction is fairly easy for physicists to understand.  One
considers the $n$ dimensional Clifford-Dirac algebra $[\gamma^a ,
\gamma^b ]_+ = 2 \delta^{ab}$ in its irreducible Dirac spinor
representation.  Normalize the trace of the unit operator in this
representation to $1$.  The trace of any projector is then a
rational number between $0$ and $1$, and it is easy to see that as
$n \rightarrow$ infinity these numbers become dense in the
interval.   Embed this sequence of finite dimensional algebras in
the algebra of all bounded operators in Hilbert space. Obviously,
each member of the sequence can be realized as a subalgebra of the
next member.  Murray and Von Neumann show that an appropriately
defined limit of this sequence is a Type $II_1$ factor, named $R$.
The factor $R$ is in fact the only $II_1$ factor which is
generated by such an increasing sequence of finite dimensional
subalgebras.  This property, with an appropriate definition of
what it means to approximate any element by a finite dimensional
matrix, is called {\it hyperfiniteness}, and is shared by its
daughter, the $II_{\infty}$ factor $R_{0,1}$.

A maximal abelian subalgebra of $R_{0,1}$ is thus generated by
finite linear combinations of orthogonal projectors, whose traces
are arbitrary positive real numbers.  If two such linear
combinations have the same collection of traces then they are
unitarily equivalent to each other.   Now consider a $9$-vector
${\bf X}$ of elements of the commuting subalgebra .  It is a limit
of terms of the form $\sum {\bf x_I} e_I$, where $e_I e_J =
\delta_{IJ} e_I$ and $Tr e_I = P^+_I $.  The subgroup of the
unitary group that preserves the maximal abelian subalgebra, acts
on these vectors by permuting the ${\bf x_I}$ with the same value
of $P^+_I$.

To whet the reader's appetite for what follows, I will consider
the Lagrangian of \cite{bfss} where both the ${\bf X}$ and
$\Theta$ variables are taken to be elements of the $R_{0,1}$
algebra rather than an $N\times N$ matrix algebra. We also drop
the variable $R$ representing the length of the compactified null
circle since we are attempting to describe the Lorentz invariant
theory with noncompact longitudinal direction. We will continue to
call elements of the Von Neumann algebra, {\it matrices}, in order
to distinguish them from quantum mechanical operators. For the
moment, we restrict attention to the maximal abelian subalgebra,
dropping the commutator terms in the Lagrangian. A general
configuration is given by

\eqn{mat1}{ {\bf X}  = \sum {\bf X}_I (t) e_I}

\eqn{mat2}{\Theta = \sum \theta_I (t) e_I .}

In terms of the ordinary variables ${\bf x}_I,\ \theta_I$ the
Lagrangian is

\eqn{matlag}{ {\cal L} = 1/2 \sum \dot{\bf x}_I^2 p_I^+ + i \theta_I
\dot{\theta}_I p_I^+ .}

This is just the light cone Lagrangian for $M$ copies of the
eleven dimensional superparticle.   When we quantize it we get the
states of $M$ supergravitons.   Configurations with permutations
of the indices, $I$ are related by unitary equivalence in the
algebra $R_{[0,1]}$ and, as in matrix theory, this is to be
treated as a gauge invariance.   This $S_M$ gauge invariance, and
the anti-commutation relations of the $\theta_I$ give us the
correct statistics of the supergravitons.  In other words, when
the Matrix Theory Lagrangian is applied to the maximal abelian
subalgebra of $R_{0,1}$, then quantization of the theory leads to
the Fock space of 11 dimensional SUGRA.  In the next section, we
will try to cast this new form of gravitational kinematics in a
manifestly Lorentz invariant form.

\section{\bf M-theory at null infinity}

The success of the AdS/CFT correspondence tempts us to construct a
manifestly covariant formalism for AFM-theory on null-infinity. This
might cause difficulties for massive particles. Null infinity is not
a manifold and the asymptotic wave-functions of massive particles
are concentrated near two of its singularities. However, at least in
eleven dimensions, all stable finite energy states of M-theory are
massless supergravitons, so a formulation on null-infinity does not
run into {\it a priori} difficulties. We will return briefly to the
question of massive particles below.

A formulation on null infinity cannot share the dynamical
properties of AdS/CFT.   This is most clearly seen in Ashtekar's
description of massless free field theory on null
infinity\cite{ashtekar}.   The coordinates of null infinity in
eleven dimensions are $(u, \Omega)$, where $u$ is null and
$\Omega$ parametrizes a 9-sphere.   This ``manifold" does not have
a metric, but only a conformal structure: the set of conformal
rescalings of the round metric on the sphere.   The coordinate $u$
is also rescaled by the conformal factor.  If $g_{ab} (\Omega )
\rightarrow \omega^2 (\Omega) g_{ab} (\Omega )$, then $ u
\rightarrow \omega u$.

The conformal group of the 9-sphere is $SO(1,10)$ and this is
interpreted as the Lorentz group of asymptotically flat
space-time.   The translation generators, in the Lorentz frame
where the metric on the sphere is round, are the vector fields
$P_{\mu} = (1, {\bf \Omega})\partial_u $,  where ${\bf \Omega}^2 =
1$, parametrizes the 9-sphere.

It is important to understand that in eleven dimensions, the
gravitational S-matrix for finite numbers of particles does not
suffer from infra-red divergences. There is no need to consider
classical gravitational radiation in the initial or final states,
and one can use the stringent asymptotic condition\cite{ashtekar},
which only allows the vacuum as an asymptotically flat solution.
Consequently there is no need to discuss the Bondi-Metzner-Sachs
group.  The definition of asymptotically flat space-time used by
relativists, allows classical gravitational radiation in all
dimensions, and the invariance group of such a formalism would
have to be the BMS group.  This is puzzling to string theorists,
who are used to computing a gravitational S-matrix with only
Poincare invariance.   The absence of IR divergences is the
explanation of this puzzle, and we will adopt the string theory
definition of asymptotically flat space, in which the classical
background is forced to satisfy the stringent asymptotic condition
that Ashtekar calls {\it restriction to the vacuum sector}.

Multiparticle states of massless particles can be described in terms
of fields at null infinity, but there are no propagation equations.
For example, a massless scalar field is completely specified by the
commutation relation \eqn{ashcom}{[ \phi (u, \Omega ), \phi (v,
\Theta ) ] = \epsilon (u - v) \delta^9 (\Omega , \Theta) ,} where we
have used the usual Heaviside $\epsilon$ function and the invariant
$\delta$ function on the sphere.  Thus, the $u$ coordinate plays a
role analogous to longitudinal position, rather than light front
time. There is no analog of light front time at null infinity.

Dynamics at null infinity is instead encoded in the S-matrix.
Indeed, so far we have only described future null infinity.  Past
null infinity is an identical copy of the same conformal
``manifold", and the scattering matrix is a mapping between the
natural bases of states on ${\cal I}_{\pm}$.   The problem of
dynamics thus reduces to finding a set of equations for
determining the scattering matrix in terms of more elementary
objects, the analogs of the Hamiltonian of the light front
formalism.

Our discussion of this problem breaks into two parts, a long
kinematical discussion of a Matrix Theory-like parametrization of
the Hilbert spaces at ${\cal I}_{\pm}$, and a short speculative
subsection on dynamical equations for the S-matrix.

\subsection{Kinematics}

The alternative to Ashtekar's Fock space description of kinematics
at null infinity is based on the work of \cite{holcosm} and
\cite{bfss}. One of the primary purposes of the present paper is to
make contact between the formalism of \cite{holcosm} and
\cite{susyholo}, and established theories of quantum gravity.

The fundamental geometrical object in Lorentzian space-time is a
causal diamond.  In the holographic proposal for the kinematic
description of quantum space-time, each causal diamond is replaced
by a Hilbert space which is the fundamental representation of the
anti-commutation relations, \eqn{acr}{[S_a (n) , S_b (m) ]_+ =
\delta_{ab} \delta_{mn}.}  The logarithm of the dimension of this
Hilbert space is the quantum version of (${1\over 2}$ times) the
area of the largest $d - 2$ surface encountered on the boundary of
the causal diamond.   This surface is called the {\it holographic
screen}.

The idea behind this association is the Cartan-Penrose relation
between pure spinors of $SO(1,10)$ and null flags consisting of a
null direction and a bit of area transverse to it.  Consider a $10$
dimensional null hypersurface in eleven dimensional space-time, and
let $n^{\mu}$ be the null direction pointing out of the surface at a
given point. Both this null direction and the orientation of the
spacelike $9$ plane orthogonal to it are captured by a pure spinor
satisfying \eqn{cp2}{n^{\mu} \gamma_{\mu} \psi = 0,} the
Cartan-Penrose equation.  Indeed, $n^{\mu} \propto \bar{\psi}
\gamma^{\mu} \psi$, and the orientation of a transverse $9$ plane is
specified by the non-vanishing components of $\bar{\psi}
\gamma^{[\mu_1 \ldots \mu_k ]} \psi$.   The pure spinor has $16$
independent real components $S_a (\bomega )$ .  $\bomega$ is a
coordinate on the holographic screen and the notation indicates that
we should think of the collection of spinor variables describing
bits of the screen as sections of the spinor bundle over the screen.
Note that in ordinary Lorentzian geometry, specifying all of these
variables for every screen would over determine the conformal
structure of the manifold.  That is, there must be consistency
conditions, relating the $S_a (\bomega )$ variables for different
screens.

The CP equation is conformally invariant, and invariant under
rescaling of $\psi$, as well as under rotations in the transverse
plane.  In classical Lorentzian geometry, the spinors are only
sensitive to the causal structure.  We think of the $S_a (n)$
operators above as the quantization of the screen variables
$S_a(\bomega )$.

   Each $S_a (n)$ should be thought of as representing a
particular pixel of the holographic screen, with a quantized area
equal (in Planck units) to $4 {\rm ln} 256$ (the logarithm of the
dimension of the irreducible representation of the Clifford algebra
with $16$ generators. Actually, since the algebra \ref{acr} is
invariant under orthogonal transformations $S_a (n) \rightarrow
O_{nm} S_a (m)$, only one basis for the $S_a (n)$ algebra should be
a associated with a pixel (a small area element on the holographic
screen).  A more invariant way to describe the pixelation of the
screen geometry is to imagine that the algebra of continuous (or
measurable) functions on the screen is replaced by a finite
dimensional algebra, with a particular basis of operators
corresponding (morally, as one says ) to operators $\int\ d\bomega
S_a (\omega ) f_n (\bomega ) ,$ with $f_n$ some basis of the finite
function algebra.

For general space-times there will not be a canonical algebra of
functions for a given causal diamond.  Rather, there are many
choices, related by the analog of general coordinate
transformations.   However, in asymptotically flat space-times it
is reasonable to insist on causal diamonds which preserve the full
symmetry of spatial rotations.   A nested sequence of causal
diamonds corresponds to a time-like observer, and it is reasonable
to insist on keeping the maximal symmetry of such an observer's
world line in the quantum theory.   For M-theory in 11 dimensions,
a possible procedure is to equate the function algebra of the
holographic screen of a causal diamond, with the matrix algebra
generated by a sequence of $d_N$ dimensional representations of
the $SO(10)$ Dirac algebra,with $d_N \rightarrow \infty$.  The
$S_a (n)$ for a nested sequence of causal diamonds, converging to
null infinity, should transform as sections of (appropriately
defined) spinor bundles over this sequence of algebras.

When the space-time boundary conditions admit a TCP
transformation, we expect to be able to choose TCP invariant
causal diamonds.  That is, we expect two descriptions of the
Hilbert space $S_a^{+} (n) = T ^{a,n}_{b,m} S_b^- (m) T^{-1}$,
related by an anti-unitary involution, $T$.  The matrix $C$
implements the geometrical space inversion symmetry as well as
charge conjugation.  We may view these two collections of
operators as being associated with the future and past null
boundary of the causal diamond respectively. As such, we expect
them to also be related by the unitary time evolution operator

\eqn{smat}{S_a^{+} (n) = S_D^{\dagger} S_a^{-} (n) S_D , } where
$S_D$ is the scattering matrix of the causal diamond.  If we view an
observer as a nested sequence of causal diamonds, the S-matrix of
one diamond can be constructed by concatenating the evolution
operators in individual diamonds, as in \cite{holcosm}. It is clear
that these observer dependent quantities cannot be exactly gauge
invariant observables of a theory of quantum gravity. However, in an
asymptotically flat space-time, the S-matrix for the limit of large
causal diamonds should be universal and gauge invariant.

Thus we want to imagine a limit in which our causal diamond
becomes the interior of null infinity in asymptotically flat 11
dimensional space-time.   In order to achieve that we first study
the single particle states of supergravitons.  These are described
by the null momentum of the state, tensored with a spin index. The
wave function has the form $\Phi_A (\bomega , p)$, where $A$ takes
on 256 values. The null momentum (in an appropriate Lorentz frame)
is $p^{\mu} = p (1, \bomega)$, and $\bomega^2 = 1$ parametrizes a
point on $S^9$. The spin space is a representation of a single
copy of the $16$ generator Clifford algebra, with generators
$S_a$.  These are the independent components of a pure spinor
satisfying $p_{\mu} \gamma^{\mu} \psi = 0$.

We can organize the operator algebra on the single particle Hilbert
space in the following way:  for any measurable section $f^a
(\bomega )$ of the spinor bundle over the $9$-sphere introduce the
operator $S (f)$ by \eqn{spinop}{S(f) \Phi_A (p, \bomega ) =
f^a(\bomega ) (S_a)_A^B \Phi_B (p, \bomega ). } These operators,
combined with the $SO(10)$ rotations of the $9$-sphere, generate the
full operator algebra for fixed value of $p$.   We can describe this
in the following language. The algebra of measurable functions on
the sphere has an outer automorphism group $SO(10)$ which preserves
the round metric on the sphere.   The operators $S (f)$ are a linear
map from sections of the spinor bundle to the algebra of operators
on Hilbert space. If we write \eqn{spinop2}{S (f) = \int d\bomega
f^a(\bomega ) S_a (\bomega ),} then, under $SO(10)$, $S_a (\bomega
)$ transforms like a section of the spinor bundle on the 9-sphere.

So far we have worked at fixed $p$, and restricted attention to
single particle states.  We deal with these omissions simultaneously
by combining Ashtekar's insight that $p$ is analogous to
longitudinal momentum on a light front, with the treatment of
multi-particle states in Matrix Theory and the theory of Type II Von
Neumann algebras.  Define the Von-Neumann algebra \eqn{alg}{{\cal A}
\equiv R_{[0,1]} \otimes {\cal M} ,} the tensor product of the
hyperfinite type $II_{\infty}$ factor with the algebra of measurable
functions on the $9$-sphere. We define the ${\cal S}$, the spinor
bundle over this algebra to be the tensor product of $R_{[0,1]}$
with the spinor bundle over the nine sphere. Introduce linear maps
$S(\rho (f) )$ for $ f \in {\cal S}$ , from ${\cal S}$ to the
algebra of operators in Hilbert space. $\rho$ is any element of a
group of outer automorphisms of ${\cal A}$, which we specify below.
We require $S$ to be invariant under inner automorphisms of ${\cal
A}$. A general element of ${\cal A}$ can be written as a limit of a
finite sum $r_i \chi_i$, where the $\chi_i$ are characteristic
functions of disjoint subsets of $S^9$, and $r_i \in R_{[0,1]}$.   A
general inner automorphism has the form $\sum U_i \chi_i$, where the
$\chi_i$ form a partition of unity on $S^9$. Thus, we can use inner
automorphism invariance to diagonalize all the $r_i$.

Let us verify that this prescription generates the Fock space of
eleven dimensional SUGRA.    Invariance under unitary
transformations means that we can write every element of the algebra
as a limit of finite sums of the form $A = \sum e_k f_k (\bomega )$,
where each $e_k$ is a projector in $R_{[0,1]}$, and $e_k e_l =
\delta_{kl} e_k$. This description is redundant, as unitary
transformations can permute projectors if they have the same trace.
However, as in Matrix Theory, this is the gauge transformation of
particle statistics. We will adopt the conventional treatment of
this symmetry in first quantized theories: we work in a large
Hilbert space on which this symmetry acts, and impose invariance
under it as a condition on states. Let $p_k = {\rm Tr}\ e_k$.  In
$R_{[0,1]}$ a projector is characterized up to inner automorphism by
its trace, so \eqn{spinop3}{S (e_k \otimes f_k) = S(p_k , f_k) . }
Linearity of $S_a$ implies that the algebra of operators $S_a (p_k ,
\bomega_k )$ combined with their images under outer automorphisms,
generate the entire operator algebra on the Hilbert space.   We
could choose the $S_a (p_k, \bomega_k )$ to anti-commute for
different values of $k$, but it is more convenient to do a Klein
transformation so that they commute.

We conclude that the Hilbert space of our system is a direct sum
of $K$ particle sectors, where $K$ is any positive integer.  $K$
corresponds to the number of independent projectors in the tensor
decomposition of an element $A \in {\cal A}$.   The $K$ particle
sector is the symmetrized tensor product of single particle
sectors.  Each single particle sector is characterized by a
positive real number $p_k$ .  For each section of the spinor
bundle on the sphere, $f^a (\bomega )$, we have an operator $S[f]
= \sqrt{p_k} S_a f^a (\bomega )$, where $[S_a, S_b ]_+  =
\delta_{ab}$.

{}From a physicist's point of view, the operators $S [f]$ are a
complete set of operators in the single particle Hilbert space. That
is, we will momentarily define operations on the $S[f] $ which
change the values of $p_k$ and $\bomega$,  by rotations and Lorentz
boosts. The infinitesimal generators will act on $S[f]$ as linear
differential operators and we will be able to write these operations
as the result of commutation with bilinears in the $S [f]$.   From
the mathematical point of view, in which we think of the $S (f)$ as
linear maps from the spinor bundle over ${\cal A}$ to the quantum
operator algebra, the full operator algebra is generated by
composing these linear operators with outer automorphisms of ${\cal
A}$.  The whole system is viewed as arising as a limit of a similar
construction for finite dimensional algebras ${\cal A}_N$,
corresponding to finite causal diamonds.   In the finite dimensional
case, all the automorphisms will be inner, and the $S_a (n)$ really
generate the operator algebra even in the strict mathematical sense.

 The variables of our system are thus concisely characterized as
operators $S(\psi )$ where $\psi$ is a section of the spinor bundle
over the algebra $R_{0,1} \otimes L_1 (S^9)$, and $S(\psi )$ is
invariant under inner automorphism of the algebra. These operators
have commutation relations \eqn{acspinop}{[ S(\psi ), S(\phi )]_+ =
(\psi , \phi ),} where the scalar product in the spinor bundle
includes the scalar product on the algebra, defined by its trace.
The map from the spinor bundle to operators is linear. To be more
explicit, a general element of the spinor bundle is $\psi = \sum A_i
\psi_i (\bomega )$, where $A_i$ is an element of $R_{0,1}$ and $\psi
(\bomega )$ a section of the ordinary spinor bundle over the
$9$-sphere. $S(\psi )$ is defined to be invariant under the inner
automorphisms $A_i \rightarrow U_i^{\dagger} A_i U_i$, where the
$U_i$ are unitary elements of $R_{0,1}$, as well as under Hermitian
conjugation in $R_{0,1}$. The scalar product is defined as
\eqn{scalprod2}{(\psi , \phi ) = \sum {\rm Tr}(A_i B_i ) \int d^9
\bomega\ \psi_i^a (\bomega ) \phi_i^a (\bomega ) .}

We have shown that the irreducible representation of this operator
algebra is precisely the Fock space of eleven dimensional
supergravitons.

I would like to emphasize the insight that the above construction
provides, regarding the operator algebra $S_a (n)$ of a finite
causal diamond.   I have already emphasized that for some choice of
the basis of labels $n$, we should think of the operator $S_a (n)$
as representing the information stored in a pixel of the holographic
screen.   Our current discussion emphasizes that this ``local"
operator algebra is the SUSY algebra of a massless superparticle.
That is, the information content in a pixel can be encoded in the
spin states of a massless supermultiplet.   This is a much clearer
statement of the relation between SUSY and holography than the one I
presented in \cite{susyholo}.

In eleven dimensions, SUSic kinematics forces us to consider a
theory of gravitation. In lower dimensions, we can have massless
supermultiplets which do not include the graviton, and the necessity
for gravitation in this holographic theory of space-time, may become
evident only at the dynamical level. The consistency conditions on
the dynamics of overlapping causal diamonds\cite{holcosm} are
discrete analogs of general coordinate invariance. At the moment,
there is no direct proof that this requires us to use
representations of the lower dimensional SUSY algebras with spin
two, but if the formalism does have a correspondence limit with low
energy effective field theory then this must be the case.

 We now want to describe how the super Poincare
algebra of 11D SUGRA acts on the operator algebra. It is sufficient
to define the action on the single particle operators $S (e_i
\otimes f )$ and then extend it to the full operator algebra by
linearity of $S(\psi )$. A general Lorentz transformation is the
product of an $SO(10)$ rotation and a boost, $B(\bomega^{\prime},
\zeta)$  along some direction, $\bomega^{\prime}$, with rapidity
$\zeta$.  The combined action is a general conformal transformation
of the nine sphere. In particular, the conformal transformation
corresponding to the boost is \eqn{lorconf}{\bomega \rightarrow
{{(\bomega\cdot \bomega^{\prime}) \bomega^{\prime} e^{\zeta} +
(\bomega^{} - (\bomega^{\prime} \cdot\bomega^{})
\bomega^{\prime})}\over \sqrt{[e^{2\zeta} -
1][\bomega^{\prime}\cdot\bomega^{\prime}]^2  + 1}}.}

For any pair of directions $\bomega$ and $\bomega^{\prime}$, let $K
(\bomega^{\prime} , \bomega )$ be the counterclockwise rotation in
the $\bomega , \bomega^{\prime} $ plane, which takes $\bomega$ into
$\bomega^{\prime} $.  If $\Lambda$ is a general element of the
conformal group $SO(1,10)$, then \eqn{little}{L (\Lambda, \bomega )
\equiv K^{-1} ( \Lambda (\bomega ), \bomega ) \Lambda ,} is in the
little group of the point $\bomega $.  Thus it is the product of a
boost $$ B(\bomega , \zeta (C,\bomega ) $$ and an $SO(9)$ rotation,
$R(C,\bomega )$ in the 9 plane perpendicular to $\bomega$ in
$R^{10}$.    We define the action of the Lorentz group of the spinor
bundle over $S^9$ as follows.   First we relate the bases of the
spinor spaces at two points by parallel transport via the rotation
$K$ \eqn{partrans}{K (\bomega^{\prime} , \bomega ) f_a (\bomega ) =
f_a (\bomega^{\prime} ),} for any section $f_a$ of the bundle.  Now
for a general conformal transformation $\Lambda$ in $SO(1,10)$ we
write \eqn{trans}{\Lambda = K(\bomega^{\prime} , \bomega )
L(\Lambda,\bomega) ,} so that the action of $\Lambda$ on the spinor
bundle is induced by the action of the little group of a point. The
latter is defined by \eqn{trans2}{L(\Lambda , \bomega ) f_a (\bomega
) = e^{{1\over 2}\zeta (\Lambda , \bomega )} D_{ab} [R(\Lambda ,
\bomega )] f_b (\bomega) ,} where $D_{ab} [R]$ is the usual sixteen
dimensional spinor representation of $SO(9)$.   With this
transformation law, the Conformal Killing Spinor Equation,
\eqn{cks}{D_{m} ^{ab} q_b^{\alpha} (\bomega ) \equiv (\partial_m
\delta_{ab} - \omega_m^{ab} \gamma_{ab}) q_b^{\alpha} = {1\over 9}
e_m^A (\gamma_A)^{ab} {\cal D}^{bc} q_c^{\alpha } ,} is Lorentz
covariant.   ${\cal D}$ is the Dirac operator on $S^9$.  $\alpha $
labels the $32$ linearly independent solutions of this equation,
which transform as a spinor under $SO(1,10)$.

We define the SUSY generators by \eqn{susy}{Q^{\alpha} \equiv S[q]
\equiv \sum \int d^9 \bomega_{(i)}\ S_a^{(i)} (\bomega_{(i)} )\
q_a^{\alpha} (\bomega_{(i)})  ,} where the sum has $K$ terms in the
$K$ supergraviton sector. These operators satisfy
\eqn{susycr}{[\bar{Q}^{\alpha} , Q^{\beta} ]_+ = \sum p_i
\bar{q}_a^{\alpha}(\bomega_{(i)} )\ q_a^{\beta} (\bomega_{(i)}) .}
The $32 \times 32$ matrix $\bar{q}_a^{\alpha}(\bomega_{(i)} )\
q_a^{\beta} (\bomega_{(i)})$, on the right hand side of this
equation can be expanded in anti-symmetric products of the 11
dimensional Dirac matrices $ \Gamma^{\mu_1 \ldots \mu_n}$, where $n
=0, 1,2$ or $5$. The transformation properties of sections of the
spinor bundle under $SO(9)$ rotations perpendicular to $\bomega_i$
show that only the matrices $1, (\Omega_{(i)})_I \Gamma^I$ and
$(\Omega_{(i)})_I \Gamma^{0I}$ are allowed.  The transformation
under boosts in the $\bomega_i$ direction allows only the linear
combination $1 + (\Omega_{(i)})_J \Gamma^J $. Thus, the right hand
side is just the momentum operator $P_{\mu}$ for positive energy
incoming particles, dotted into $\Gamma^{\mu}$ . Our Hilbert space
carries a representation of the 11 dimensional super-translation
algebra.

We now define the action of the Lorentz group on the operator
algebra by \eqn{transS}{U^{\dagger} S(e_i \otimes f )U(\Lambda ) =
S(\rho (\bomega, \Lambda ) [e_i] \otimes f^{(\Lambda )}).} Here
$f^{(\Lambda )}$ is the transformed element of the spinor bundle
over the sphere, which we defined above.  $\rho(\Lambda , \bomega )$
is an element of the automorphism group of $R_[0,1]$ .  Recall that
$\lambda_{\rho}$ is defined by \eqn{trdef}{{\rm Tr}\ \rho[a] =
e^{\lambda_{\rho}} {\rm Tr}\ a,} for every element of the algebra.
Let $J(\Lambda , \bomega )$ be the Jacobian of the conformal
transformation $\Lambda$ on the nine sphere.   If we choose
\eqn{fixaut}{e^{\lambda_{\rho} (\Lambda , \bomega )} = J^{-1}
e^{-\zeta (\Lambda , \Omega )},} (recall that $\zeta$ is the
rapidity of the boost in the little group) then the anti-commutation
relations \eqn{acrinv}{[S (\psi ) , S(\phi )]_+ = (\psi , \phi ),}
are invariant under the action of the Lorentz group, as they must be
if this action is implemented by a unitary transformation.

Conversely, because our Hilbert space is defined as the irreducible
representation of the anti-commutation relations, there is a unitary
action, unique up to a multiple of the identity, which implements
the Lorentz group.   Thus we have described the full action of the
super-Poincare algebra on our Hilbert space.  This is not a big
surprise, since we have already identified this space as the Fock
space of supergravitons.   Nonetheless it is interesting to see the
role of the automorphism group of the algebra in the explicit
construction.   I believe that these formulae for the action of the
super-Poincare algebra will be useful in the attempt to construct
the S-matrix.   I now turn to a brief discussion of that, as yet
unrealized, program.

\subsection{Dynamics}

The prehistory of string theory was the search for an alternative
method for constructing a scattering matrix consistent with
unitarity, Lorentz invariance, and causality.   Field theory gave
such a construction, but left an enormous amount of ambiguity. Today
we recognize that ambiguity as the existence of many possible fixed
points of the renormalization group.   That is to say, the ambiguity
is connected to the high energy behavior of the theory.

This is very explicit in the S-matrix theorist's derivation of
unitarity.   One starts with an assumed spectrum of particles and
makes an asymptotic expansion of the S-matrix \eqn{sexp}{S = 1 + i
\sum_{n=1}^{\infty} T_n .} The unitarity condition becomes
\eqn{uni}{T_n - T_n^{\dagger} = \sum_{k = 1}^{n-1} T_k T_{n-k} .}
Assuming appropriate analyticity conditions (presumed to follow from
causality) for S-matrix elements, one then claims that this relation
determines the amplitudes in terms of $T_1$. It explicitly computes
discontinuities across cuts in terms of low order amplitudes.
Dispersion relations (Cauchy's theorem ) should then enable us to
compute the full amplitude.  $T_1$ itself is Hermitian and, if there
are a finite number of particles, analyticity shows that it can be
written in terms of an integral over a local, Lorentz scalar,
Lagrangian.   The problem is that the Cauchy integrals do not
converge until we take some momentum derivatives, which leads to
polynomial ambiguities in higher order amplitudes.   These are what
field theorists call {\it renormalization counterterms}, and the
procedure suffers precisely the ambiguities of the local field
theory with the same tree level Lagrangian.

String theory was born as an attempt to find a different and more
unique solution by positing an infinite number of stable particles
(at zeroth order) so that $T_1$ (the Veneziano-Virasoro-Shapiro-
Koba-Nielsen amplitudes) could have \lq\lq better" high energy
behavior, corresponding to a series of Regge poles.  We all know the
story of how this inadvertently led to the construction of a theory
of quantum gravity, and most of us also know that tree level string
theory does not give a correct description of the high energy
behavior of quantum gravity amplitudes.   Rather, it is believed
that the generic regime of large kinematic invariants is dominated
by the production and decay of black holes\cite{tbwfdlgt} .

A possible route to the construction of the scattering matrix for
11D SUGRA then, would be to follow the old S-matrix program, using
the added input of supersymmetry and insights about black hole
dominance of high energy interactions.   A starting point might be
the ideas of 't Hooft\cite{thooft} .   More particularly, consider
an $N$ particle scattering amplitude as a function of the following
three variables: the center of mass energy$^2$, $s$, the subenergy
$s_{n-1}$ of a cluster of $n - 1$ of the particles, and the impact
parameter $b$ between the $n$th particle and the cluster.   In the
limit $s \sim s_{n -1} \gg M_P $ and $b \gg R_S (s_{n - 1}) $, (the
Schwarzschild radius of the cluster), the following approximation to
the $n$ particle S-matrix suggests itself: $S_n = S_{n-1}
e^{i\delta}$ The first factor is the exact $n - 1$ particle
S-matrix, while $e^{i\delta}$ is the scattering amplitude of a
single particle in the classical field of a black hole of mass $s_{n
-1}$ (in the Lorentz frame where the total spatial momentum of the
cluster vanishes).  Perhaps this information about high energy
behavior will help to determine the amplitudes\footnote{We know
other things about inclusive cross sections for black hole
production and decay in certain kinematical regimes, which might be
useful.}

The form of the SUSY generators in our formalism, suggests an
entirely different approach to finding the scattering matrix. Note
that, although our formula has non-vanishing values for all
components of these generators, the contribution of any single
particle state has only $16$ non-vanishing components. This is the
familiar fact that the supergraviton representation of 11D SUSY is
BPS.   Note further that if we write the algebra corresponding to
outgoing rather than incoming  particles, then we find \lq\lq the
other half" of the components of each particle's supergenerators.

This is reminiscent of Matrix Theory, in light cone frame, where
half of the SUSY generators are kinematical.   The other half have a
non-trivial, but fairly simple, construction in terms of kinematical
particle variables, {\it which encodes the entire dynamics of the
theory}.   By analogy one should seek a formula which expresses the
incoming components of the SUSY generators as simple functions of
the outgoing particle variables $S(f)$.   Since incoming and
outgoing variables are related by the S-matrix, this formula would,
be a constraint on the S-matrix; perhaps enough of a constraint to
determine it.

Finally, recall that he general approach to holographic space-time
presented in \cite{holcosm} contains a consistency condition which
is very hard to satisfy.   Namely, a quantum space-time is defined
in terms of Hilbert spaces and time evolution operators for causal
diamonds, plus a system of overlap conditions designed to guarantee
that two observers who share a piece of their respective causal
diamonds, describe that piece in a consistent manner.   This
condition is very hard to implement and so far the only successful
example corresponds to a particular space-time, the FRW universe
with $p=\rho$.   We conjectured that this consistency condition
would completely determine the dynamics of possible theories of
quantum gravity.

In asymptotically flat space, one would like to translate the
consistency conditions on local causal diamonds, into constraints on
the S matrix.   Recall that the S matrix is the common limit of an
infinite set of sequences of local S-matrices for particular
observers.   So far, I have not been able to translate the dynamical
compatibility conditions into equations for the S-matrix, but this
should be possible.

It should be clear to the reader that these are just ideas for
ideas.   The problem of finding a non-perturbative formulation of
string/M-theory for general asymptotically flat space-times is
unsolved, and would appear to be the most important unsolved formal
problem in the subject.

\section{Massive particles and compactification}

Two different issues arise when trying to generalize these
considerations to situations with fewer non-compact dimensions
and/or less SUSY.  The first is the necessity of describing stable
massive particles in a formalism based on null infinity.   The
second is the appearance of non-gravitational multiplets in systems
with less than maximal SUSY.

For massive BPS states the problems of working at null infinity seem
tractable.  Indeed, our formalism is not really local at null
infinity since we work at fixed longitudinal momentum. It is easy to
describe massive BPS states in terms of an extended momentum space
including the central charges.   Already at the level of finite
causal diamonds, we simply replace the pixel algebra by the SUSY
algebra with appropriate central charges.
 However, we know of examples of stable massive states in string
theory which are not BPS\footnote{The simplest is the spinor in
$SO(32)$ heterotic string theory, and the K-theory classification of
D branes gives rise to other examples. It is a reasonable hypothesis
that all stable massive particles in asymptotically flat string
theory, carry a K theory charge, which will be a torsion element for
non-BPS particles.}.

It should be noted that in general, the masses of even BPS particles
is not determined by symmetries.   The masses will appear in the
pixel algebra, as part of the kinematics, and will have to be
determined by the dynamical equations of the theory.

A prescription for incorporating K theory charges into the operator
algebra of a holographic screen is also the key to understanding
compactification in this formalism.   The K theory charges of states
are the only topological features of the internal manifold that are
preserved in string theory, since ordinary topology is not invariant
under duality transformations. Thus, except in certain limits, one
should not be able to think of the compact dimensions of space-time
in terms of ordinary geometrical notions.  By contrast, our
formalism implies a rather direct relation between geometrical
notions in the non-compact dimensions and the structure of the
quantum theory.   The quantum formalism for finite causal diamonds
has a causal structure, which determines that of the Lorentzian
geometry that emerges in the large area limit.   The conformal
factor of that Lorentzian geometry is directly related to the size
of Hilbert spaces.  By contrast, duality invariant information about
the geometry of the compact space is incorporated in the pixel
algebra of the non-compact space.

As an aside, we should note that de Sitter space should be
considered non-compact in the sense in which the phrase is used
here.   The global dS manifold can be foliated by compact spatial
sections, but the holographic formalism describes only the static
observer's horizon volume.   The observer's cosmological horizon
converges to null infinity in the large radius limit, and the
interesting physics of the system is concerned with the way in which
this limit is approached.   The global manifold is really only a
trick (the thermofield doubling trick) for discussing the thermal
physics of a single horizon volume.

\section{\bf Conclusions}

I have introduced the von Neumann algebra $R_{0,1}$ in an attempt to
construct a non-perturbative Lorentz invariant formulation of the
quantum theory underlying 11 dimensional supergravity. The
manifestly covariant kinematics on null infinity uses $R_{0,1}$ to
incorporate the Matrix Theory description of multi-particle states.
So far however it is only a kinematics.

One of the most significant results in this paper is the deeper
understanding we have achieved of the relation between SUSY and
holographic screens.   The Cartan-Penrose equation gives us a way
to relate the variables describing a pixel on a holographic screen
to a pure spinor.   The commutation relations of the pure spinor
variables are identical to those of the reduced SUSY algebra for a
massless superparticle with fixed momentum.   In eleven dimensions
this implies that  the quantum states of a holographic pixel are
precisely the spin states of the SUGRA multiplet.

I have attempted to make contact between a holographic formulation
of quantum space-time, and extant descriptions of certain
space-times in terms of string/M-theory.   The holographic
formulation is maximally local: its key ingredient is the operator
algebra of a causal diamond, which should be thought of as a
quantization of Cartan-Penrose variables belonging to the (dual
space of the) spinor bundle of the diamond's holographic screen. The
geometrical information encoded in these variables are the
orientation of a pixel on the screen, as well as its area. An
observer following a time-like trajectory is modeled as a nested
sequence of causal diamonds. A quantum space-time is a topological
spatial lattice with such a quantum observer attached to each point,
together with overlap/consistency conditions that enforce agreement
between the joint observations of different observers.   It is very
hard to find solutions of these conditions, so perhaps they are the
only dynamical information the formalism needs.   Indeed, the single
known solution\cite{holcosm} automatically describes the dynamical
evolution of a flat $p =\rho$ FRW cosmology.

The local formulation is gauge dependent and is formulated in
generic (unitary) gauge.   That it must be so follows from the
principle of general covariance, but also from the more profound
holographic principle.   Indeed the holographic principle (in the
form advocated by Fischler and the present author) implies that a
finite area causal diamond has a finite set of observables and
therefore cannot make infinitely precise measurements of itself.
This means that local physics is intrinsically ambiguous.   The
claim is that this ambiguity is the quantum origin of general
coordinate invariance.   Gauge invariant formulations of
gravitational systems exist only when space-time has an infinite
asymptotic boundary.   In this paper I tried to show how the
conventional S-matrix description of 11D asymptotically flat
space-time could be obtained as a limit of the local formulation of
holographic space-time.   The basic idea was to take the function
algebras of a sequence of causal diamonds to be an increasing
sequence of representations of the $SO(10)$ Clifford-Dirac algebra,
converging to the algebra $R_{0,1} \otimes {\cal M} (S^9 )$: the
tensor product of the hyperfinite $II_{\infty}$ factor, and the
algebra of measurable functions on the nine sphere.   I showed that
if the quantum algebra of observables was the space of linear
functionals $S(a)$ on the spinor bundle of this algebra, invariant
under inner automorphisms, then the obvious limit of the finite
anti-commutation relations gave us precisely the Fock space of $11D$
SUGRA.

Although I did not go into detail, I also presented the basic idea
for incorporating compactification in this framework.  The nine
sphere is replaced by a lower dimensional sphere, and the algebra
incorporates charges encoding the quantum numbers of finite energy
wrapped branes on the compact manifold.   Questions remain about
torsion elements in the K-theory classification of D-branes.   In
general, at the kinematic level, it seems that one will have to
include the masses of particles as parameters, and hope that they
will be determined by the dynamical equations.

These equations themselves remain a mystery.   I presented several
directions of research for determining them.   Perhaps, in the
future, someone will pay attention to these ideas and figure them
out.   I wish her well.
\section{Acknowledgments}

I would like to thank Willy Fischler, Lorenzo Mannelli and Tomeu
Fiol for contributing to the suite of ideas that formed the basis of
this paper. This research was supported in part by DOE grant number
DE-FG03-92ER40689.




  %




\end{document}